\documentclass[twocolumn,preprintnumbers,
nofootinbib,prd,floatfix]{revtex4}

\usepackage{graphicx}

\usepackage[hypertex]{hyperref}
\newcommand{\beq}{\begin{equation}}
\newcommand{\eeq}{\end{equation}}

\begin{document}

\pagestyle{plain}

\preprint{MADPH-08-1508 \hspace{.1cm} \\ MAD-TH-08-06}

\title{Deflected Mirage Mediation: A 
Framework for Generalized Supersymmetry Breaking}

\author{Lisa~L.~Everett, Ian-Woo Kim, Peter Ouyang, and Kathryn M.~Zurek}
\affiliation{Department of Physics, University of Wisconsin,
Madison, WI 53706, USA}


\begin{abstract}
We present a general phenomenological framework for dialing between gravity mediation, gauge mediation, and anomaly mediation.  The approach is motivated from recent developments in moduli stabilization, which suggest that gravity mediated terms can be effectively loop suppressed and thus comparable to gauge and anomaly mediated terms. The gauginos exhibit a mirage unification behavior at a ``deflected" scale, and gluinos are often the lightest colored sparticles.  The approach provides a rich setting in which to explore generalized supersymmetry breaking at the LHC.

\end{abstract}
\maketitle

High energy physics is currently at the dawn of a new era, in which the Large Hadron Collider (LHC) will
test many ideas for physics beyond the Standard Model (SM).  Of these ideas, softly broken supersymmetry (see \cite{reviews} for reviews) is a strong contender for new TeV scale physics.   The phenomenology of such theories, including the minimal supersymmetric
standard model (MSSM), depends crucially on the superpartner mass spectrum, which is governed by the soft supersymmetry breaking sector.

In constructing models of supersymmetry breaking, it has long been understood that viable scenarios are most easily obtained if spontaneous supersymmetry breaking occurs in a hidden sector and is
communicated to the visible sector by ``mediator" fields via loop-suppressed or nonrenormalizable interactions.    The model-dependent hidden sector dynamics can be represented by supersymmetry breaking order parameters, the vacuum expectation values of the auxiliary components (``F terms" or ``D terms") of the mediators.  The visible sector supersymmetry breaking scale is given by the
order parameter divided by the mediation scale (up to loop factors). Standard cases are gravity mediation \cite{gravmed}, gauge mediation \cite{gaugemed}, and
anomaly mediation \cite{anomalymed}. It is often assumed that one mediation mechanism dominates, usually to address a given phenomenological problem of the MSSM.

In this letter, we develop a general framework for dialing continuously between these three different scenarios of supersymmetry breaking (gravity/moduli, gauge and anomaly mediation), focusing on the fully mixed case, when all three types contribute comparably to the soft masses.   By taking the correct limit, any single mediation mechanism, or combination of any two, can be recovered.  Indeed, a single mediation mechanism is often not dominant.  This occurs in the class of string models with stabilized moduli proposed by  Kachru, Kallosh, Linde, and Trivedi (KKLT) \cite{Kachru:2003aw}.  In the KKLT scenario, the scale of the moduli mediated terms is suppressed by $1/\ln{(M_P/m_{3/2}})$ ($M_P$ is the reduced Planck mass and $m_{3/2}$ is the gravitino mass)
 \cite{Choi:2004sx}.   Phenomenological models based on the KKLT construction thus have comparable gravity mediated and anomaly mediated soft terms, solving the negative slepton mass problem of anomaly mediation.
 This mixed scenario is known as ``mirage mediation,"  and it has been studied extensively \cite{miragepheno,littlehierarchy,Choi:2007ka}.

We show here that  it is natural in mirage mediated models to have comparable contributions to visible sector supersymmetry breaking from gauge mediation.  For example, a given string model usually contains light vectorlike exotics as well as moduli, leading to gauge mediated contributions once local supersymmetry is broken.  This leads to what we call ``deflected mirage mediation," in which gravity/moduli, gauge, and anomaly mediation all play important roles in supersymmetry breaking.  The gauge mediation contribution introduces threshold effects that deflect the renormalization group flow of the soft masses, which also solve the negative slepton mass problem of anomaly mediation (this aspect is similar to deflected anomaly mediation  \cite{Pomarol:1999ie}).
While this framework is applicable to other contexts than the fully mixed case of deflected mirage mediation, the richest and most novel phenomenology occurs in the fully mixed limit.  

Let us explicitly demonstrate that in a broad class of effective supergravity models (which might be
realizable in string theory) the contributions to the supersymmetry breaking terms from gravity mediation, gauge mediation, and anomaly mediation are comparable.   For supergravity models with a
modulus field $T$ as the mediator, the gravitino mass is $m_{3/2} \sim F^T/M_P$, while the observable sector supersymmetry breaking scale
$m_{\rm soft}$ is
\begin{equation}
\label{graveq}
m_{\rm soft}^{({\rm grav})} \sim  \frac{F^T}{T+\overline{T}}.
\end{equation}
In both anomaly and gauge mediation, the interactions between the mediators and the SM fields are loop suppressed.   For anomaly mediation, the mediator is the supergravity conformal compensator field $C$, and the gravitino mass is of the order $m_{3/2}\sim F^C/C$.  The mediator $X$ of gauge mediation is a SM singlet  which couples to messenger fields; the gravitino mass is $\sim F^X/M_P$.  In each case, the visible sector soft terms are
\begin{equation}
\label{anomgaugeq}
m_{\rm soft}^{({\rm anom})}\sim \frac{1}{16\pi^2} \frac{F^C}{C},\;m_{\rm soft}^{({\rm gauge})}\sim \frac{1}{16\pi^2} \frac{F^X}{X}.
\end{equation}
To see why these scales are of similar sizes in deflected mirage mediation, recall the mirage unification scenario motivated from KKLT.  The effective supergravity theory for $T$ is given by the K\"{a}hler potential $K=K_0-3\ln(T+\overline{T})$ and superpotential $W=W_0-\mathcal{A}e^{-aT},$ ($\mathcal{A}\sim M_P^3$, $a$ is a numerical factor due to gaugino condensation, and $K_0,\,W_0$ stabilize other moduli), and the uplifting potential $V_{\rm uplift} \sim (T+\overline{T})^{-n_p}$.
$T$ is stabilized to $aT \simeq \ln{M_P/m_{3/2}}\sim 4\pi^2$ \cite{Kachru:2003aw,Choi:2004sx}, such that
\begin{equation}
\label{mirageresult}
\frac{F^T}{T+\overline{T}}\simeq \frac{1}{aT}m_{3/2} \sim  \frac{1}{4\pi^2} \frac{F^C}{C}.
\end{equation}
Let us now consider the presence of $X$ and the $N$ messenger pairs $\Psi_{1,2}$. The K\"{a}hler potential for $X$ (with modular weight $n_X$)  is $(T+\overline{T})^{-n_X} X\overline{X}$.
To see heuristically
 that $F^X/X\sim F^C/C$ in a broad class of models, recall that in supergravity, the F term of $X$ (neglecting mixing) is
\begin{equation}
\label{fdef}
F^X \simeq -e^{\frac{K}{2}}K^{X\overline{X}}D_{\overline{X}}\overline{W}=-e^{\frac{K}{2}}K^{X\overline{X}}(\partial_{\overline{X}}\overline{W}+(\partial_{\overline{X}}K) \overline{W}),
\end{equation}
where $K^{X\overline{X}}$ is the inverse K\"{a}hler metric.  If $X$ has no bare superpotential mass term (in a given string model, $X$ would be a massless string mode) but has self-couplings at higher order, then $\partial_{\overline{X}}\overline{W}=0$ at the minimum, and 
\begin{eqnarray}
F^X &\simeq &-e^{\frac{K}{2}}K^{X\overline{X}}(\partial_{\overline{X}}K) \overline{W}\\
&=&
-(e^{\frac{K}{2}}\overline{W})(T+\overline{T})^{n_X}(T+\overline{T})^{-n_X}X =-m_{3/2}X, \nonumber
\end{eqnarray}
such that
\begin{equation}
\label{heuristicarg}
\frac{F^X}{X}\simeq -\frac{F^C}{C}.
\end{equation}
Eq.~(\ref{heuristicarg}) holds for 
$\partial_{\overline{X}}\overline{W} \sim (\partial_{\overline{X}}K) \overline{W}$, and for a general class of K\"{a}hler potentials $K=f(XX^\dagger)$. 
We have found that for a general superpotential with high scale $\Lambda$:
\begin{equation}
W=W_0-\mathcal{A}e^{-aT}+\frac{X^n}{\Lambda^{n-3}}+\lambda X\Psi_1\Psi_2,
\end{equation}
keeping only ${\cal O}(X^{2n-2})$, ${\cal O}(m_{3/2}X^n)$, and ${\cal O}(m^2_{3/2}X^2)$ terms (dropping subleading terms of order $X/M_P$, $X/\Lambda$, or $1/\ln(M_P/m_{3/2})$), $F^X/X$ is given by (details will be given in a forthcoming publication \cite{uslong}):
\begin{equation}
\label{genarg}
\frac{F^X}{X}\simeq -\frac{2}{n-1}\frac{F^C}{C}.
\end{equation}
Despite the presence of $T$, the result
is identical to that of deflected anomaly mediation
\cite{Pomarol:1999ie};  it is obtained even if the K\"{a}hler
potential of $T$ is not of the precise ``no-scale" form but
instead is $-p\ln(T+\overline{T})$.  Furthermore, this result holds for a generic K\"{a}hler potential for $X$ as long as $\langle X  \rangle \ll M_{\rm P}$.  The relative negative
sign between $F^X/X$ and $F^C/C$
can change if $X$ appears in the superpotential with a
negative power due to strong dynamics.

The effective tree-level supergravity theory of the observable sector has the gauge kinetic function $f_a=T$, the matter field K\"{a}hler potential $K_{\rm obs}=(T+\overline{T})^{-n_i}\Phi_i\overline{\Phi}_i$ ($n_i$ is the modular weight of the MSSM field $\Phi_i$), and the superpotential $W_{\rm obs}= y^0_{ijk}\Phi_i\Phi_j\Phi_k$ ($y^0_{ijk}$ are (unnormalized) Yukawas, assumed to be independent of $X$ and $T$.  We assume bare superpotential mass parameters are absent but otherwise do not address the $\mu$ problem; instead,  $\mu$ and $B$ are replaced by $m_Z$ and $\tan\beta$ after electroweak symmetry breaking.  The theory includes $N$ vectorlike messenger pairs which lie in complete $SU(5)$ multiplets, to maintain gauge coupling unification.  

The MSSM soft supersymmetry breaking terms at the high scale $M_{\rm G}\sim 2\times 10^{16}$ GeV and the threshold effects at the messenger scale $M_{\rm mess}= \langle X \rangle$ are (here $M_{\rm mess}^+$, $M_{\rm mess}^-$ denote scales just above and below $M_{\rm mess}$):\\

\noindent $\bullet$ {\bf Gaugino masses}:
\begin{eqnarray}
\label{gaugino}
  M_a (M_{\rm G}) &=& \frac{F^T}{T+\overline{T}}
  + \frac{g_0^2}{16\pi^2} b'_a \frac{F^C}{C} \\
  M_a (M_{\rm mess}^-) &=& M_a(M_{\rm mess}^+)
+ \Delta M_a,
\end{eqnarray}
in which
\begin{eqnarray}
  \Delta M_a = - N\frac{g_a^2(M_{\rm mess})}{16\pi^2} \left( \frac{F^C}{C}
  + \frac{F^X}{X} \right).
   \label{gauginothresh}
\end{eqnarray}
$g_0$ is the unified gauge coupling at $M_{\rm G}$, and the $b'_a$ are the gauge coupling beta function coefficients.
Our convention is  $b'_a<0$ for asymptotically free theories.\\

\noindent $\bullet$ {\bf Trilinear scalar couplings}:
\begin{eqnarray}
A_{ijk} = A_i+ A_j + A_k,
\end{eqnarray}
in which
\begin{eqnarray}
\label{trilinear}
A_i (M_{\rm G}) &=& (1-n_i) \frac{F^T}{T+\overline{T}}
- \frac{\gamma_i}{16\pi^2} \frac{F^C}{C},
\end{eqnarray}
$\gamma_i = 2 \sum_a g_a^2 c_a(\Phi_i) - \sum_{lm} |y_{ilm}|^2/2$ ($y_{ijk}$ are normalized Yukawas and $c_a$ is the quadratic Casimir).  There are no messenger-induced threshold effects for the trilinears. \\

\noindent $\bullet$ {\bf Soft scalar mass-squared parameters}:
\begin{eqnarray}
&&m_i^2 (M_{\rm G})= (1-n_i) \left|\frac{F^T}{T+\overline{T}}\right|^2\\
&&-\frac{\theta'_i}{32 \pi^2} \left( \frac{F^T}{T+\overline{T}}
\frac{F^{\overline{C}}}{\overline{C}} + h.c. \right ) - \frac{\dot{\gamma}'_i}{(16\pi^2)^2} \left|\frac{F^C}{C}\right|^2 \nonumber
\end{eqnarray}
\begin{equation}
m_i^2 ( M_{\rm mess}^- ) = m_i^2 (M_{\rm mess}^+)
+ \Delta m_i^2
\end{equation}
in which
\begin{eqnarray}
\label{msqthr}
\Delta m_i^2 = \sum_a 2 c_a N
\frac{g_a^4 (M_{\rm mess}) }{(16\pi^2)^2}
\left | \frac{F^X}{X} + \frac{F^C}{C} \right |^2.
\end{eqnarray}
In the above, $\dot{\gamma}_i=2\sum_a g_a^4b_a c_a(\Phi_i) - \sum_{lm} |y_{ilm}|^2b_{y_{ilm}}$,
$\theta_i=4 \sum_a g_a^2 c_a(\Phi_i) - \sum_{lm} |y_{ilm}|^2(3-n_i-n_l-n_m)$.   The primed quantities include the messenger contributions, while the unprimed quantities take on MSSM values. 

The anomaly and gauge mediated terms can be expressed in terms of $m_0\equiv m_{\rm soft}^{({\rm grav})}$ (see Eq.~(\ref{graveq})).  As in \cite{miragepheno}, we define $\alpha_{\rm m}$ by $F^C/C=\alpha_{\rm m} \ln(M_P/m_{3/2})m_0$ ($\alpha_{\rm m}\equiv \alpha$ in mirage mediation), and $\alpha_{\rm g}$ by $F^X/X = \alpha_{\rm g} F^C/C$. The standard KKLT model is recovered for $\alpha_{\rm m}=1$ and $N=0$.  For $\alpha_{\rm m}, \alpha_{\rm g}$ of order 1, the gravity, anomaly, and gauge mediation terms are all comparable. 
The supersymmetric flavor problem can be alleviated with universal modular weights, though if the gluinos and stops are light, the FCNC constraints must be reassessed. The supersymmetric CP problem is present if the phases of the F terms result in irremovable relative phases in the soft terms.   In mirage mediation, the  F terms can be taken to be real without loss of generality \cite{miragepheno}; whether this holds in deflected mirage mediation 
 depends on the stabilization mechanism for $X$.  Here we assume real F terms and defer a detailed study of CP violation to future work.
\begin{figure}
\includegraphics[height=2.8in]{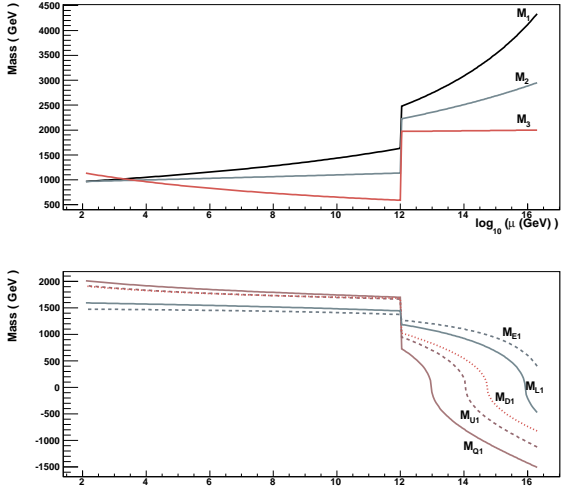}
\caption{The renormalization group evolution of the gaugino masses (top panel) and the soft scalar masses (bottom panel) of the first generation for point A, in which $M_{\rm mess}=10^{12}\mbox{GeV}$.    For the scalar masses, $M_{fi}\equiv m^2_{fi}/\sqrt{|m^2_{fi}|}$.}
\label{pointArg}
\end{figure}

In describing the patterns of soft terms, we consider the ``deflected mirage'' case, where all mediation mechanism contributions are roughly the same size.  To show the types of effects which can appear, we take the fiducial case $\alpha_{\rm m}=1$.
The gauge piece enters through the threshold effects at the messenger scale, which can shift the gaugino masses up or down;  the phenomenologically more interesting case occurs when the gaugino masses are lowered.  The gaugino masses still unify at a mirage scale.
By dialing the relative sizes of the terms in Eq.~(\ref{gauginothresh}), the gaugino mirage unification scale can be easily tuned to a desired value (in contrast to the standard KKLT scenario)
as in deflected anomaly mediation \cite{Rattazzi:1999qg}.  Lowering the gaugino mass through these threshold effects can also lead to quasi-conformal renormalization group running of the soft scalar masses, as shown in Fig.~\ref{pointArg}.
\begin{table}
\caption{The MSSM particle mass spectrum for points A and B. All masses are in GeV. }
\begin{tabular}{cc|cc||cc|cc}
\hline
\hline
\multicolumn{4}{c||}{Point A} & \multicolumn{4}{c}{Point B} \\
\hline
$h$ & 117  & $H,A$ & 1529  &  $h$ & 116  & $H,A$ & 865  \\
$\tilde{g}$ & 1170 & $H^\pm$ & 1531 & $\tilde{g}$ & 1130 & $H^\pm$ & 869 \\
$\chi^0_1$ & 1003 & $\chi^0_2$ & 1015 & $\chi^0_1$ & 608 & $\chi^0_2$ & 683 \\
$\chi^0_3$ & 1374 & $\chi^0_4$ & 1380 & $\chi^0_3$ & 818 & $\chi^0_4$ & 844 \\
$\chi^\pm_1$ & 1011 & $\chi^\pm_2$ & 1369 & $\chi^\pm_1$ & 682 & $\chi^\pm_2$ & 835 \\
$\tilde{u}_L$ & 1965 & $\tilde{u}_R$ & 1890 & $\tilde{u}_L$ & 1164 & $\tilde{u}_R$ & 1140 \\
$\tilde{d}_L$ & 1974 & $\tilde{d}_R$ & 1888 & $\tilde{d}_L$ & 1172 & $\tilde{d}_R$ & 1148 \\
$\tilde{e}_L$ & 1587  & $\tilde{e}_R$ & 1470  & $\tilde{e}_L$ & 783  & $\tilde{e}_R$ & 709  \\
$\tilde{\mu}_L$ & 1587  & $\tilde{\mu}_R$ & 1470  & $\tilde{\mu}_L$ & 783  & $\tilde{\mu}_R$ & 709  \\
$\tilde{t}_1$ & 1420 & $\tilde{t}_2$ & 1791 & $\tilde{t}_1$ & 860 & $\tilde{t}_2$ & 1113 \\
$\tilde{b}_1$ & 1769 & $\tilde{b}_2$ & 1872 & $\tilde{b}_1$ & 1059 & $\tilde{b}_2$ & 1141 \\
 $\tilde{\tau}_1$ & 1459 & $\tilde{\tau}_2$ & 1583 & $\tilde{\tau}_1$ & 702 & $\tilde{\tau}_2$ & 782 \\
\hline
\hline
\end{tabular}
\end{table}
\begin{figure}
\includegraphics[height=2.8in]{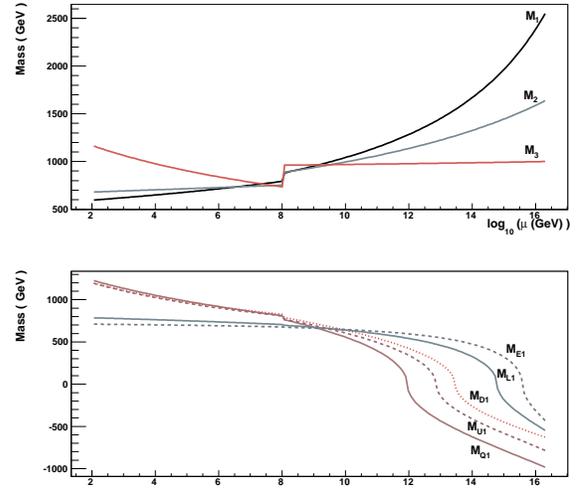}
\caption{The renormalization group evolution of the gaugino masses (top panel) and the first generation soft scalar masses (bottom panel) for Point B, in which $M_{\rm mess}=10^8\mbox{GeV}$.  }
\label{pointBrg}
\end{figure}

In  Fig.~\ref{pointArg} and Table~1, we show the renormalization group evolution and mass spectrum for our sample point A, in which $m_0=2 \mbox{ TeV}$, $N=3$, $\tan\beta=10$, ${\rm sign}(\mu)=+1$,  $n_i=1$ for the Higgs doublets and $n_i=1/2$ for all other fields, 
$\alpha_{\rm g}=1$  ($F^C/C = F^X/X \sim 65 \mbox{ TeV}$),  and $M_{\rm
mess} = 10^{12} \mbox{ GeV}$.  The gaugino mirage scale is 
\begin{equation}
M_{\rm mirage} = M_{\rm G}\left (\frac{m_{3/2}}{M_P} \right )^{\alpha_{\rm m}\rho/2},
\end{equation}
in which 
\begin{equation}
\rho= \frac{1+ \frac{2Ng_0^2}{16\pi^2}  \ln \frac{M_{\rm GUT}}{M_{\rm mess}}}{1- \frac{\alpha_{\rm m} \alpha_{\rm g} Ng_0^2}{16\pi^2} 
 \ln \frac{M_P}{m_{3/2}}}.
\end{equation}
 The low energy gaugino mass ratios  (see e.g.~\cite{Choi:2007ka}) are  $M_1:M_2:M_3\sim 1:1:1.2$.
The scalar masses are deflected from mirage unification, as shown in Fig.~\ref{pointArg}.  If $M_{\rm mess}$ is sufficiently high, then the light generation scalars can exhibit near-mirage unification below $M_{\rm mess}$.  Yukawa effects further spoil the mirage behavior  for the third generation.  If the gaugino mirage unification scale is of order a TeV, a mixed bino-wino  \cite{ArkaniHamed:2006mb} lightest supersymmetric partner (LSP) results, which in this case results in an experimentally allowed dark matter relic abundance (the bino-wino mass difference is small and the higgsinos are heavy due to the large $\mu$ term is needed for electroweak symmetry breaking). 

In Fig.~\ref{pointBrg} and Table~1, we show sample point B, in which $\alpha_{\rm m}=1$, $\alpha_{\rm g}=-1/2$, $m_0=1$ TeV, $M_{\rm mess}=10^8  \mbox{ GeV}$, and the other parameters as in point A.   The messenger scale is now below the standard mirage unification scale of $\sim 10^{10} \mbox{GeV}$.  The low energy gaugino mass ratios are $\sim 1:1.1:2$.  The spectrum is relatively compressed, with light sleptons and a light stop; these features are also present in gauge messenger models \cite{gaugemessenger}.   In this case, the lighter stau is the next-to-lightest superpartner (NLSP).  Lighter gluinos and stops thus appear naturally in this mixed limit, such that the LHC phenomenology will focus on gluino and stop production (see \cite{uslong}).

Another limit is to switch off anomaly mediation and consider comparable gauge and gravity mediated terms ($\alpha_{\rm m} \rightarrow 0$, $\alpha_{\rm g}\rightarrow \infty$, $\alpha_{\rm m}\alpha_{\rm g}$ finite).    
This results in a stretched low energy spectrum with lighter gauginos than scalars, such that 
the LHC phenomenology is dominated solely by gluino production.  For further details, see \cite{uslong}.

Mirage mediation with a TeV mirage scale is known to ameliorate the little hierarchy problem \cite{littlehierarchy}. 
Here it is easy to obtain a TeV gaugino mirage unification scale, but the fine-tuning is not significantly reduced, as the scalars do not unify at this scale and no new A terms (needed for Higgs mass corrections \cite{Dermisek:2006ey}) are present.   

In conclusion, we have provided a method for dialing continuously between 
gravity, gauge, and anomaly mediated supersymmetry breaking.   Motivated by the KKLT framework
for moduli stabilization, 
we showed that comparable gauge-mediated terms can also be present in a broad class of models.    This deflected mirage mediation scenario exhibits rich features, including light gauginos and light stops.  Our method can be used as a general 
framework for studying the collider and cosmological implications of TeV scale supersymmetry.  As the LHC era nears, it is of particular importance to extend studies of low energy supersymmetry beyond standard scenarios in which a single mediation mechanism dominates.  

Note added: in the final stages of preparing this paper, a similar scenario was presented in \cite{Nakamura:2008ey} emphasizing the cosmological moduli problem and the $\mu/B\mu$ problem.



\acknowledgments
This work is supported by the U.S. Department of Energy grant DE-FG-02-95ER40896.  P.O. is also supported by the NSF Career Award PHY-0348093 and a Cottrell Scholar Award. We thank K.~Choi, 
D.~Demir, 
G.~Kane,  
and G.~Shiu for helpful comments.



\begin{thebibliography}{99}

\bibitem{reviews}
  S.~P.~Martin,
  hep-ph/9709356; 
  D.~J.~Chung,  L.~Everett, G.~Kane, S.~King, J.~Lykken and L.~T.~Wang,
  Phys.\ Rept.\  {\bf 407}, 1 (2005)
  [hep-ph/0312378].

 \bibitem{gravmed}
  A.~H.~Chamseddine, R.~Arnowitt and P.~Nath,
  Phys.\ Rev.\ Lett.\  {\bf 49}, 970 (1982);
  R.~Barbieri, S.~Ferrara and C.~Savoy,
  Phys.\ Lett.\  B {\bf 119}, 343 (1982);
  L.~Hall, J.~Lykken and S.~Weinberg,
  Phys.\ Rev.\  D {\bf 27}, 2359 (1983).

 \bibitem{gaugemed}
  M.~Dine, A.~Nelson and Y.~Shirman,
  Phys.\ Rev.\  D {\bf 51}, 1362 (1995)
  [hep-ph/9408384];
  M.~Dine, A.~Nelson, Y.~Nir and Y.~Shirman,
  Phys.\ Rev.\  D {\bf 53}, 2658 (1996)
  [hep-ph/9507378]; 
  G.~Giudice and R.~Rattazzi,
  Phys.\ Rept.\  {\bf 322}, 419 (1999)
  [hep-ph/9801271].

\bibitem{anomalymed}
  L.~Randall and R.~Sundrum,
  Nucl.\ Phys.\  B {\bf 557}, 79 (1999)
  [hep-th/9810155];
  G.~Giudice, M.~Luty, H.~Murayama and R.~Rattazzi,
  JHEP {\bf 9812}, 027 (1998)
  [hep-ph/9810442]; 
  J.~Bagger, T.~Moroi and E.~Poppitz,
  JHEP {\bf 0004}, 009 (2000)
  [hep-th/9911029].


\bibitem{Kachru:2003aw}
  S.~Kachru, R.~Kallosh, A.~Linde and S.~P.~Trivedi,
  Phys.\ Rev.\  D {\bf 68}, 046005 (2003)
  [hep-th/0301240].

 \bibitem{Choi:2004sx}
  K.~Choi, A.~Falkowski, H.~Nilles, M.~Olechowski,  S.~Pokorski,
  JHEP {\bf 0411}, 076 (2004)
  [hep-th/0411066]; 
  K.~Choi, A.~Falkowski, H.~Nilles and M.~Olechowski,
  Nucl.\ Phys.\  B {\bf 718}, 113 (2005)
  [hep-th/0503216].

\bibitem{miragepheno}
  K.~Choi, K.~Jeong and K.~Okumura,
  JHEP {\bf 0509}, 039 (2005)
  [hep-ph/0504037];
M.~Endo, M.~Yamaguchi and K.~Yoshioka,
  Phys.\ Rev.\  D {\bf 72}, 015004 (2005)
  [hep-ph/0504036]; A.~Falkowski, O.~Lebedev and Y.~Mambrini,
  JHEP {\bf 0511}, 034 (2005)
  [hep-ph/0507110]; H.~Baer, E.~Park, X.~Tata and T.~Wang,
  JHEP {\bf 0608}, 041 (2006)
  [hep-ph/0604253], JHEP {\bf 0706}, 033 (2007)
  [hep-ph/0703024].
  
  \bibitem{littlehierarchy}
  K.~Choi, K.~Jeong, T.~Kobayashi, and K.~Okumura,
  Phys.\ Lett.\  B {\bf 633}, 355 (2006)
  [hep-ph/0508029], 
  Phys.\ Rev.\  D {\bf 75}, 095012 (2007)
  [hep-ph/0612258];
  R.~Kitano and Y.~Nomura,
  Phys.\ Lett.\  B {\bf 631}, 58 (2005)
  [hep-ph/0509039];
  O.~Lebedev, H.~Nilles and M.~Ratz,
  hep-ph/0511320;  
  A.~Pierce and J.~Thaler,
  JHEP {\bf 0609}, 017 (2006)
  [hep-ph/0604192].


\bibitem{Choi:2007ka}
  K.~Choi and H.~Nilles,
  JHEP {\bf 0704}, 006 (2007)
  [hep-ph/0702146].
  
  
\bibitem{Pomarol:1999ie}
  A.~Pomarol and R.~Rattazzi,
  JHEP {\bf 9905}, 013 (1999)
  [hep-ph/9903448].

 \bibitem{uslong}
  L.~Everett, I.-W.~Kim, P.~Ouyang, K.~Zurek, to appear.
  
  
  \bibitem{Rattazzi:1999qg}
  R.~Rattazzi, A.~Strumia and J.~D.~Wells,
  Nucl.\ Phys.\  B {\bf 576}, 3 (2000)
  [hep-ph/9912390].
  
  \bibitem{ArkaniHamed:2006mb}
  N.~Arkani-Hamed, A.~Delgado and G.~F.~Giudice,
  Nucl.\ Phys.\  B {\bf 741}, 108 (2006)
  [hep-ph/0601041].
  
  \bibitem{gaugemessenger}
  R.~Dermisek, H.~D.~Kim and I.~W.~Kim,
  JHEP {\bf 0610}, 001 (2006)
  [hep-ph/0607169];
  K.~J.~Bae, R.~Dermisek, H.~D.~Kim and I.~W.~Kim,
  JCAP {\bf 0708}, 014 (2007)
  [hep-ph/0702041].
  
  \bibitem{Dermisek:2006ey}
  R.~Dermisek and H.~D.~Kim,
  Phys.\ Rev.\ Lett.\  {\bf 96}, 211803 (2006)
  [hep-ph/0601036].

\bibitem{Nakamura:2008ey}
  S.~Nakamura, K.~Okumura and M.~Yamaguchi,
 arXiv:0803.3725 [hep-ph].

\end{thebibliography}
\end{document}